\documentstyle[preprint,aps]{revtex}
\newcommand{\sect}[1]{\setcounter{equation}{0}\section{#1}}
\newcommand{\bfm}[1]{\mbox{\boldmath${#1}$}}
\renewcommand{\theequation}{\arabic{section}.\arabic{equation}}
\newcommand{\app}{\setcounter{section}{0}
\setcounter{equation}{0} \renewcommand{\thesection}{APPENDIX
\Alph{section}}\renewcommand{\theequation}{\Alph{section}.\arabic{equation}}}
\begin{document}
\draft
\title{Cole-Hopf Like Transformation for Schr\"odinger
Equations Containing Complex Nonlinearities}
\author{G. Kaniadakis, and A.M. Scarfone\footnote{\small e-mail:
kaniadakis@polito.it, scarfone@polito.it}}
\address{ Dipartimento di Fisica - Politecnico di Torino -
Corso Duca degli Abruzzi 24, 10129 Torino, Italy; \\
Istituto Nazionale di Fisica della Materia - Unit\'a del
 Politecnico di Torino, - Italy}
\date{\today}
\maketitle
\begin {abstract}
We consider systems, which conserve the particle number and are
described by Schr\"odinger equations containing complex
nonlinearities. In the case of canonical systems, we study their
main symmetries and conservation laws. We introduce a Cole-Hopf
like transformation both for canonical and noncanonical systems,
which changes the evolution equation into another one containing
purely real nonlinearities, and reduces the continuity equation
to the standard form of the linear theory. This approach allows us
to treat, in a unifying scheme, a wide variety of canonical and
noncanonical nonlinear systems, some of them already known in the
literature.\\ 
\end {abstract}
\sect{Introduction} Over the last few decades many nonlinear
Schr\"odinger equations (NLSEs) have been proposed in order to
test the fundamental postulates of quantum mechanics, like for
instance, the Bialynicki-Birula and Mycielski equation
\cite{Bialyniki}, the Kostin equation \cite{Kostin}, the Gisin
equation \cite{Gisin} among many others \cite{Dodonov}. In Ref.
\cite{Weinberg1} a wide class of NLSEs for finite-dimensional
quantum systems was selected in order to preserve the {\sl
homogeneity principle} of the original Schr\"odinger equation,
with the {\sl superposition principle} being destroyed by the
nonlinear terms.

Many of the NLSEs proposed in literature contain complex
nonlinearities. For instance, the Doebner-Goldin (DG) equations
\cite{Doebner1,Doebner2,Doebner3} were introduced as the most
general class of Schr\"odinger equations, compatible with the
Fokker-Planck equation for the probability density $\rho=|\psi|^2$
namely $\partial\,\rho/\partial\,t+{\bfm\nabla}\cdot{\bfm
j}_0=D\,\Delta\,\rho$, being ${\bfm j}_0$ the standard quantum
current and $D$ a positive diffusion coefficient. The importance
of this class of evolution equations is that it is founded on the
grounds of the group theory: the nonlinear terms were derived from
the representation analysis of the $Diff({\bfm R}^3)$ group which
was proposed as a {\sl universal quantum kinematical group}
\cite{Goldin}.

In addition, a large number of NLSEs with complex nonlinearities
have been proposed in order to describe some phenomenologies in
condensed matter physics. For instance, in Ref. \cite{Gagnon} a
NLSE with a nonlinearity of the type
$a_1\,|\psi|^2\,\psi+a_2\,|\psi|^4\,\psi+
i\,a_3\,\partial_x(|\psi|^2\,\psi)+(a_4+i\,a_5)\,\partial_x|\psi|^2\,\psi$
is introduced to describe a single mode wave propagation in a Kerr
dielectric guide. Another example is given by the generalized
Ginsburg-Landau equation appeared in Ref. \cite{Petviashvili}.
This equation contains the nonlinearity:
$a_1\,|\psi|^2\,\psi+i\,a_2\,\psi+
i\,a_3\,\partial_{xx}\psi+i\,a_4\,|\psi|^2\,\psi$ \cite{Malomed1}
which takes into account pumping and dumping effects of the
nonlinear media and can be used to describe dynamical modes of
plasma physics, hydrodynamics and also, solitons in optical fibers
(Ref. \cite{Malomed2} and references therein). Finally, complex
nonlinearities in Schr\"odinger equations are used to describe
propagation of high power optical pulses in ultrashort soliton
communication systems \cite{Doktorov,Facao}, incoherent solitons,
\cite{Bang1,Bang2}, multi-channel bit-parallel-wavelength optical
fiber networks \cite{Kivshar}, among others.

In this paper we consider the most general class of NLSEs
conserving the quantity $N=\int|\psi|^2\,d^nx$:
\begin{eqnarray}
i\,\hbar\,\frac{\partial\,\psi}{\partial\,t}=-\frac{\hbar^2}{2\,m}\,\Delta\,\psi+(W+i\,{\cal
W})\,\psi \ ,\label{sch1}
\end{eqnarray}
where the real $W$ and imaginary $\cal W$ parts of the complex
nonlinearity are smooth functions of the fields $\psi,\,\psi^\ast$
and their spatial derivatives of any order. When $\psi$ is written
in polar representation $\psi=\rho^{1/2}\,\exp(i\,S/\hbar)$, Eq.
(\ref{sch1}) is split into two nonlinear partial differential
equations for the real fields $\rho$ and $S$:
\begin{eqnarray}
&&\frac{\partial\,\rho}{\partial\,t}+{\bfm\nabla}\cdot\left(\frac{{\bfm\nabla}\,S}{m}\,\rho+{\bfm
F}\right)=0 \ ,\label{hjcb}\\
&&\frac{\partial\,S}{\partial\,t}+\frac{({\bfm\nabla}\,S)^2}{2\,m}+W+U_q=0
\ ,\label{hjca}
\end{eqnarray}
where:
\begin{eqnarray}
U_q=-\frac{\hbar^2}{4\,m}\,\frac{\Delta\,\sqrt{\rho}}{\sqrt{\rho}}
\ ,
\end{eqnarray}
is the quantum potential \cite{Bhom} and the real functional $\bfm
F$ is related to $\cal W$ through
\begin{eqnarray}
{\cal W}={\hbar\over2\,\rho}\,{\bfm\nabla}\cdot{\bfm F} \ ,
\end{eqnarray}
as the particle number conservation requires. It is easy to
recognize that Eq. (\ref{hjcb}) is a nonlinear continuity
equation, which involves only the term $\cal W$, while Eq.
(\ref{hjca}) is a nonlinear {\sl Hamilton-Jacobi} like equation
involving only the term $W$.\\ In the Calogero picture
\cite{Calogero}, the system (\ref{hjcb}), (\ref{hjca}) is
$C$-integrable if there exists a transformation of the dependent
or/and independent variables: $t\rightarrow T,\,\,{\bfm
x}\rightarrow{\bfm X},\,\,\rho\rightarrow R,\,\,S\rightarrow{\cal
S}$ which transforms Eqs. (\ref{hjcb}), (\ref{hjca}) into:
\begin{eqnarray}
&&\frac{\partial\,R}{\partial\,T}+
\overline{\bfm\nabla}\cdot\left(\frac{\overline{\bfm\nabla}\,{\cal
S}}{m}\,R\right)=0 \ ,\label{hjc1b}\\ &&\frac{\partial\,{\cal S}
}{\partial\,T}+\frac{(\overline{\bfm\nabla}\,{\cal S})^2
}{2\,m}+\overline{U}_q=0 \ ,\label{hjc1a}
\end{eqnarray}
$\overline{\bfm\nabla}$ and $\overline{U}_q$ being the gradient
and the quantum potential in the new variables. Eqs.
(\ref{hjc1b}), (\ref{hjc1a}) constitute the well known
hydrodynamic representation of the standard linear Schr\"odinger
equation.

The principal aim of this paper is to introduce a nonlinear
transformation for the field $S$: $S\rightarrow{\cal S}$, in
order to reduce Eq. (\ref{hjcb}) to the standard form of the
linear theory (\ref{hjc1b}). As a consequence of this
transformation, the evolution equation  (\ref{sch1}) transforms
into another one containing a purely real nonlinearity. Moreover,
the current, when expressed in terms of the new field
$\phi=\rho^{1/2}\,\exp(i\,{\cal S}/\hbar)$, reduces to the
standard bilinear form of the linear Schr\"odinger theory.

The paper is organized as follows. In Sect. II, we introduce a
general class of $(n+1)$ canonical NLSEs, invariant over the
action of the $U(1)$ group. In Sect. III, starting from the
Noether theorem, we consider the main symmetries and related
conserved quantities of the canonical system. In Sect. IV, we
introduce a Cole-Hopf like transformation which eliminates the
imaginary part of the nonlinearity in the evolution equation,
while in Sect. V, the same transformation is considered in the
case of a noncanonical systems. In Sect. VI, in the framework of
the approach developed in the previous sections, we treat, in a
unifying context, some NLSEs already known in the literature, in
order to show that all the transformations introduced by the
various authors to study these equations can be obtained as
particular cases of the transformation here proposed. Finally,
some conclusions and remarks are reported in Sect. VII.
\sect{The canonical model} Let us consider the class of canonical
NLSEs described by the Lagrangian density:
\begin{eqnarray}
{\cal
L}=i\,\frac{\hbar}{2}\,\left(\psi^\ast\,\frac{\partial\,\psi}{\partial\,t}-
\psi\,\frac{\partial\,\psi^\ast}{\partial\,t}\right)-
\frac{\hbar^2}{2\,m}\,|{\bfm\nabla}\,\psi|^2-U[\psi^\ast,\,\psi] \
,\label{hamilton}
\end{eqnarray}
where ${\bfm\nabla}\equiv(\partial_1,\,\cdots,\,\partial_n)$ is
the $n$-dimensional gradient operator. The last term in the r.h.s.
of Eq. (\ref{hamilton}) is the nonlinear potential which we assume
to be a real smooth function of the fields $\psi$ and $\psi^\ast$
and their spatial derivatives. Here and in the following, we use
the notation $U[a]$ to indicate the dependence of $U$ on the field
$a$ and its spatial derivative of any order. We deal with
dynamical systems described by Eq. (\ref{hamilton}) which are
invariant under transformations belonging to $U(1)$ group. As we
will show in the next section, this condition imposes a constraint
on the form of the nonlinear potential $U$.

We start from the action
\begin{eqnarray}
{\cal A}=\int{\cal
L}\,d^nx\,dt \ ,\label{action}
\end{eqnarray}
and observe that the evolution equation of the field $\psi$ is
given by:
\begin{eqnarray}
\frac{\delta\,{\cal A}}{\delta\,\psi^\ast}=0 \ .\label{vv}
\end{eqnarray}
The functional derivative is defined through \cite{Olver}:
\begin{eqnarray}
\frac{\delta}{\delta\,a}\int{\cal
G}[a]\,d^nx=\sum_{[k=0]}(-1)^k{\cal D}_{_{I_k}}\left[
\frac{\partial\,{\cal G}[a]}{\partial\,({\cal
D}_{_{I_k}}a)}\right] \ ,\label{derfun}
\end{eqnarray}
with ${\cal D}_{_{I_k}}\equiv\partial^k/(\partial
x_{_1}^{i_1}\cdots x_{_n}^{i_n})$ and
$\sum_{[k=0]}\equiv\sum_{k=0}^\infty\,\sum_{I_k}$. The sum
$\sum_{I_k}$ is over the multi-index
$I_k\equiv(i_1,\,i_2,\,\cdots,\,i_n)$ where $1\leq p\leq n$,
$0\leq i_p\leq k$ and $\sum i_p=k$.\\ Eq. (\ref{vv}) assumes the
form:
\begin{eqnarray}
\nonumber &&\frac{\delta}{\delta\,\psi^\ast}\int
i\,\frac{\hbar}{2}\left(\psi^\ast\,\frac{\partial\,\psi}{\partial\,t}-
\psi\,\frac{\partial\,\psi^\ast}{\partial\,t}\right)\,d^nx\,dt=
\frac{\delta}{\delta\,\psi^\ast}\int\frac{\hbar^2}{2\,m}\,|{\bfm\nabla}\,\psi|^2\,d^nx\,dt+
\frac{\delta}{\delta\,\psi^\ast}\int U[\psi^\ast,\,\psi]\,d^nx\,dt
\ ,\\
\end{eqnarray}
which, after performing the functional derivatives, transforms to
the following NLSE:
\begin{eqnarray}
i\,\hbar\,\frac{\partial\,\psi}{\partial\,t}=-\frac{\hbar^2}{2\,m}\,\Delta\,\psi
+\frac{\delta}{\delta\,\psi^\ast}\int
U[\psi^\ast,\,\psi]\,d^nx\,dt \ ,\label{schroedinger}
\end{eqnarray}
where $\Delta\equiv\partial^{^2}_1+\cdots+\partial^{^2}_n$ is the
Laplacian operator. Eq. (\ref{schroedinger}) can finally be
written in the form:
\begin{eqnarray}
i\,\hbar\,\frac{\partial\,\psi}{\partial\,t}=-\frac{\hbar^2}{2\,m}\,\Delta\,\psi+\Lambda[\rho,\,S]\,\psi
\ ,\label{schroedinger1}
\end{eqnarray}
where the complex nonlinearity
\begin{eqnarray}
\Lambda[\rho,\,S]=W[\rho,\,S]+i\,{\cal W}[\rho,\,S] \
,\label{lambda}
\end{eqnarray}
has real $W[\rho,\,S]$ and imaginary ${\cal W}[\rho,\,S]$ part
defined through:
\begin{eqnarray}
&&W[\rho,\,S]=\frac{\delta}{\delta\,\rho}\int
U[\rho,\,S]\,d^nx\,dt \ ,\label{real}\\ &&{\cal
W}[\rho,\,S]=\frac{\hbar}{2\,\rho}\frac{\delta}{\delta\,S}\int
U[\rho,\,S]\,d^nx\,dt \ ,\label{imaginary}
\end{eqnarray}
$\rho$ and $S$ being the hydrodynamic fields related with the
wavefunction $\psi$ through \cite{Bhom,Madelung}:
\begin{eqnarray}
\psi({\bfm x},\,t)=\rho^{1/2}({\bfm
x},\,t)\,\exp\left[\frac{i}{\hbar}\,S({\bfm x},\,t)\right] \ .
\end{eqnarray}
\sect{Symmetries} In this section we study the main symmetries and
conserved quantities of the system described by the Lagrangian
(\ref{hamilton}).

Let us consider the $U(1)$ invariance condition. The variation
$\delta_\epsilon\,\psi=i\,\epsilon\,\psi$, with $\epsilon$ an
infinitesimal real parameter, implies the following variation on
the action:
\begin{eqnarray}
\delta_\epsilon\,{\cal
A}=-\epsilon\,\hbar\,\int\frac{\partial}{\partial\,S}U[\rho,\,S]\,d^nx\,dt
\ .\label{var1}
\end{eqnarray}
Taking into account the Noether theorem \cite{Noether} we can
also write the variation  $\delta_\epsilon\,{\cal A}$ in the form
\begin{eqnarray}
\delta_\epsilon\,{\cal
A}=-\epsilon\,\hbar\int\partial_\mu\,j_\mu[\psi^\ast,\,\psi]\,d^nx\,dt
\ .\label{var2}
\end{eqnarray}
By comparing Eq. (\ref{var1}) and (\ref{var2}) we obtain:
\begin{eqnarray}
\frac{\partial\,\rho}{\partial\,t}+{\bfm\nabla}\cdot{\bfm
j}=\frac{\partial\,U}{\partial\,S} \ .\label{continuity}
\end{eqnarray}
with $\rho=j_0$. Eq. (\ref{continuity}) is not a continuity
equation because the Lagrangian (\ref{hamilton}), for a general
nonlinear potential $U[\rho,\,S]$, is not $U(1)$-invariant. In
appendix B we show that $U(1)$-symmetry can be restored if one
assumes that the nonlinear potential $U[\rho,\,S]$ depends on $S$
only through its spatial derivative, modulo a total derivative
term, which does not change the dynamics of the system (null
Lagrangian). As a consequence, the r.h.s of Eq. (\ref{continuity})
vanishes and it becomes a continuity equation for the conserved
density $\rho$. Thus, the $U(1)$-invariance limits the class of
nonlinear potentials appearing in Eq. (\ref{hamilton}). In the
following, we consider only $U(1)$-invariant systems, where the
functional $U[\rho,\,S]$ depends on $S$ trough its spatial
derivative. For this class of systems, Eq. (\ref{continuity})
becomes:
\begin{eqnarray}
\frac{\partial\,\rho}{\partial\,t}+{\bfm\nabla}\cdot{\bfm j}=0 \
,\label{continu}
\end{eqnarray}
The conserved quantity associated to the continuity equation
(\ref{continu}) is:
\begin{eqnarray}
N=\int\rho\,d^nx \ .\label{numero}
\end{eqnarray}
The expression of $\bfm j$ is obtained in appendix A, and is given
by:
\begin{eqnarray}
j_i=\frac{\partial_i\,S}{m}\,\rho+\sum_{[k=0]}\,\frac{(-1)^k}{
f_i^{^{I_{k+1}}}}\,{\cal
D}_{_{I_k}}\left[\frac{\partial\,U[\rho,\,S]}{\partial({\cal
D}_{_{i,I_k}}S)}\right] \ ,\label{corrente1}
\end{eqnarray}
where
\begin{eqnarray}
f_i^{^{I_{k+1}}}=n-\sum_{r\not=i}^n\,\delta_{0,m_{_r}} \ ,
\end{eqnarray}
for $I_{_{k+1}}=(m_1,\,m_2,\,\cdots,\,m_n)$ with $1\leq r\leq n$,
$0\leq m_r\leq n$ and $\sum_r m_{_r}=k+1$.\\ Note that the
expression (\ref{corrente1}) of $\bfm j$, can also be written (see
appendix A) in the form:
\begin{eqnarray}
j_i=\frac{\partial_i\,S}{m}\,\rho+\frac{\delta}{\delta(\partial_iS)}\int
U[\rho,\,S]\,d^nx\,dt \ .\label{ccc}
\end{eqnarray}
Eq. (\ref{ccc}) can be obtained starting directly from the
equation (\ref{schroedinger1}) after adopting the hypothesis that
$U[\rho,\,S]$ depends on the field $S$ only through its spatial
derivatives as required from the U(1) symmetry.

In the following we consider the main space-time symmetries of the
Lagrangian (\ref{hamilton}). We note that $U[\rho,\,S]$ depends on
the variables $\bfm x$ and $t$ only trough the fields $\rho$ and
$S$, thus the system is invariant over space-time translations.
From Noether theorem we have:
\begin{eqnarray}
\frac{\partial\,{\cal
T}_\mu}{\partial\,t}+{\bfm\nabla}\cdot{\bfm{\cal T}}_{\mu}=0 \
,\label{contem}
\end{eqnarray}
where ${\cal T}_\mu\equiv T_{0\mu}$; $({\bfm{\cal T}}_\mu)_i\equiv
T_{i\,\mu}$ with $\mu=0,\cdots,3$. The components of the
energy-momentum tensor $T_{\mu\nu}$ (see appendix A) are given by:
\begin{eqnarray}
T_{00}=&&\frac{\hbar^2}{2\,m}\,|{\bfm\nabla}\,\psi|^2+U[\rho,\,S]
\ ,\label{prima}\\
T_{0j}=&&i\,\frac{\hbar}{2}\,(\psi^\ast\,\partial_j\,\psi
-\psi\,\partial_j\,\psi^\ast) \ ,\label{seconda}\\ \nonumber
T_{i0}=&&-\frac{\hbar^2}{2\,m}\left(\partial_i\psi^\ast\,\partial_t\psi-
\partial_i\psi\,\partial_t\psi^\ast\right)
\\
&&+\sum_{[k=0]}\sum_{[p=0]}^k(-1)^pB_{_{j,I_q}}^{^{I_k}}{\cal
D}_{_{I_q}}\left\{{\cal
D}_{_{I_p}}\left[\frac{\partial\,U[\rho,\,S]}{\partial\left({\cal
D}_{_{i,I_k}}\rho\right)}\right]\,\partial_t\rho+{\cal
D}_{_{I_p}}\left[\frac{\partial\,U[\rho,\,S]}{\partial\,\left({\cal
D}_{_{i,I_k}}S\right)}\right]\,\partial_tS\right\}
 \ ,\label{terza}\\ \nonumber
T_{ij}=&&-\frac{\hbar^2}{2\,m}\left(\partial_i\psi^\ast\,\partial_j\psi+
\partial_j\psi^\ast\,\partial_i\psi\right)+\delta_{ij}{\cal
L}\\&&-\sum_{[k=0]} \sum_{[p=0]}^k(-1)^pB_{_{j,I_q}}^{^{I_k}}{\cal
D}_{_{I_q}}\left\{{\cal
D}_{_{I_p}}\left[\frac{\partial\,U[\rho,\,S]}{\partial\,\left({\cal
D}_{_{i,I_k}}\rho\right)}\right]\partial_j\rho+{\cal
D}_{_{I_p}}\left[\frac{\partial\,U[\rho,\,S]}{\partial\,\left({\cal
D}_{_{i,I_k}} S\right)}\right]\partial_jS\right\} \
.\label{quarta}
\end{eqnarray}
From Eq. (\ref{prima}) and (\ref{seconda}) we obtain the conserved
quantities:
\begin{eqnarray}
&&E=\int\left[\frac{\hbar^2}{2\,m}\,|{\bfm\nabla}\,
\psi|^2+U[\rho,\,S]\right]\,d^nx\label{energia} \ ,\\ &&{\bfm
P}=-i\,\frac{\hbar}{2}\int\
\left(\psi^\ast\,{\bfm\nabla}\,\psi-\psi\,{\bfm\nabla}\,\psi^\ast\right)\,d^nx
\ ,\label{momento}
\end{eqnarray}
which are respectively the total energy and the linear momentum of
the system. We can see that $U[\rho,\,S]$ modifies the expression
of the energy while the momentum maintains the form of the linear
theory. From Eqs. (\ref{terza}) and (\ref{quarta}) we see that
the presence of $U[\rho,\,S]$ also modifies the expressions of the
fluxes associated to $E$ and $\bfm P$.

Note that if the energy-momentum tensor is symmetric in the
spatial indices $T_{ij}=T_{ji}$, the potential $U[\rho,\,S]$ is
invariant over the action of the orthogonal group $SO(n)$. In
this way we can define $n\,(n-1)/2$ conserved quantities:
\begin{eqnarray}
L^{a_1,\cdots,a_{n-2}}=\epsilon^{a_1,\cdots,a_{n-2},i,j}\int\,x_i\,T_{0j}\,d^nx
\ ,
\end{eqnarray}
where $\epsilon^{a_1,\cdots,j}$ is the $n$-rank totally
antisymmetric tensor defined as $\epsilon^{1,\cdots,1}=1$. For
$n=3$ we recognize the well known conserved components of the
angular momentum.

Finally, we look at the Galilei invariance. We recall that if the
system admits this symmetry, the corresponding generator:
\begin{eqnarray}
{\bfm G}={\bfm P}\,t-m\,N\,{\bfm x}_c \ ,\label{gener}
\end{eqnarray}
is conserved, namely:
\begin{eqnarray}
\frac{\partial\,{\bfm G}}{\partial\,t}=0 \ ,
\end{eqnarray}
where the linear momentum is given by:
\begin{eqnarray}
{\bfm P}=\int\rho\,{\bfm \nabla}S\,d^nx \ ,
\end{eqnarray}
while the mass center vector is defined as
\begin{eqnarray}
{\bfm x}_c={1\over N}\,\int\rho\,{\bfm x}\,d^nx \ .\label{mean}
\end{eqnarray}
We
consider now the Ehrenfest relation:
\begin{eqnarray}
\frac{\partial\,{\bfm x}_c}{\partial\,t}={1\over N}\,\int{\bfm
j}\,d^nx \ ,\label{ehren}
\end{eqnarray}
which can be obtained from Eq.s (\ref{continu}) and (\ref{mean}).
We note the formal similarity with the corresponding relation of
the linear theory. Here the expression of $\bfm j$ is given by Eq.
(\ref{corrente1}) and depends on the form of the nonlinear
potential $U[\rho,\,S]$. Taking into account the conservation of
$\bfm P$, and after assuming uniform conditions, from Eq.
(\ref{gener}) we obtain:
\begin{eqnarray}
\frac{\partial\,G^i}{\partial\,t}=-m\,\sum_{[k=0]}\,(-1)^k\,(f_i^{^{I_k}})^{-1}\,
\int{\cal
D}_{_{I_k}}\left[\frac{\partial\,U[\rho,\,S]}{\partial({\cal
D}_{_{i,I_k}}S)}\right]\,d^nx \ .\label{galilei}
\end{eqnarray}
Eq. (\ref{galilei}) shows that the presence of $U[\rho,\,S]$
breaks the Galilei invariance, which can be restored if ${\cal
W}=0$ as we can verify easily by using Eq. (\ref{imaginary}).
\sect{A Cole-Hopf like transformation}Let us introduce a unitary
transformation of the field $\psi$:
\begin{eqnarray}
\psi({\bfm x},\,t)\rightarrow\phi({\bfm x},\,t)={\cal
U}[\rho,\,S]\,\psi({\bfm x},\,t) \ ,\label{trasf1}
\end{eqnarray}
with ${\cal U}^\ast={\cal U}^{-1}$ so that:
\begin{eqnarray}
|\psi|^2=|\phi|^2=\rho \ .
\end{eqnarray}
The functional $\cal U$ is chosen to eliminate the imaginary part
of the NLSE (\ref{schroedinger1}) and, at the same time, to
transform the current $\bfm j$, given by Eq. (\ref{corrente1}),
into another current ${\bfm j}\rightarrow{\bfm J}$ having the
canonical form:
\begin{eqnarray}
{\bfm J}=\frac{{\bfm\nabla}\,{\cal S}}{m}\,\rho \ ,\label{corcor}
\end{eqnarray}
with $\cal S$ being the phase of the new field $\phi$:
\begin{eqnarray}
\phi({\bfm x},\,t)=\rho^{1/2}({\bfm
x},\,t)\,\exp\left[\frac{i}{\hbar}\,{\cal S}({\bfm x},\,t)\right]
\ .
\end{eqnarray}
We write ${\cal U}[\rho,\,S]$ as follows:
\begin{eqnarray}
{\cal
U}[\rho,\,S]=\exp\left(\frac{i}{\hbar}\,\sigma[\rho,\,S]\right) \
,\label{trasf2}
\end{eqnarray}
and observe that the generator $\sigma[\rho,\,S]$ is a real
functional which allows us to calculate $\cal S$ starting from
$S$:
\begin{eqnarray}
{\cal S}=S+\sigma[\rho,\,S] \ .\label{phase}
\end{eqnarray}
The generator $\sigma[\rho,\,S]$ can be obtained easily by
combining Eq.s (\ref{ccc}), (\ref{corcor}) and (\ref{phase}):
\begin{eqnarray}
\partial_i\,\sigma[\rho,\,S]=\frac{m}{\rho}\,\frac{\delta}{\delta(\partial_i\,S)}\,
\int U[\rho,\,S]\,d^nx\,dt \ .\label{gen}
\end{eqnarray}
Eq. (\ref{gen}) imposes a condition on the form of the nonlinear
potential which can be obtained using the relation:
$\partial_{ij}\,\sigma=\partial_{ji}\,\sigma$:
\begin{eqnarray}
\left[\partial_i\,\left(\frac{1}{\rho}\,\frac{\delta}{\delta(\partial_j\,S)}\right)\,
-\partial_j\,\left(\frac{1}{\rho}\,\frac{\delta}{\delta(\partial_i\,S)}\right)\right]\,
\int U[\rho,\,S]\,d^nx\,dt=0 \ .\label{rot}
\end{eqnarray}
Condition (\ref{rot}) selects the potentials $U[\rho,\,S]$ and
the nonlinear systems where we can perform the transformation
(\ref{trasf1}). In the case of one-dimensional systems the
transformation (\ref{trasf1}) is always accomplished.\\ It is now
easy to verify that the transformation (\ref{trasf1}) reduces the
evolution equation (\ref{schroedinger1}) to the following NLSE:
\begin{eqnarray}
i\,\hbar\,\frac{\partial\,\phi}{\partial\,t}=-\frac{\hbar^2}{2\,m}\,\Delta\,\phi
+\widetilde{W}[\rho,\,{\cal S}]\,\phi \ ,\label{schroedinger2}
\end{eqnarray}
which now contains only the real nonlinearity
$\widetilde{W}[\rho,\,{\cal S}]$ given by :
\begin{eqnarray}
\widetilde{W}[\rho,\,{\cal S}]=&&W+
{({\bfm\nabla}\,\sigma)^2\over2\,m}-\frac{{\bfm
J}\cdot{\bfm\nabla}\sigma}{\rho}
-\frac{\partial\,\sigma}{\partial\,t} \ ,\label{real1}
\end{eqnarray}
where $W\equiv W[\rho,\,S[\rho,\,{\cal S}]]$. The phase $\cal S$
appears in Eq. (\ref{schroedinger2}) only trough its spatial
derivatives, consequently the arbitrary integration constant,
deriving from the definition of $\cal U$, does not produce effects
and can be posed equal to zero. Note that $\widetilde W$ depends
implicitly on the field $\cal S$. In fact, Eq.
(\ref{phase}) defines $S$ as a function of $\rho$ and $\cal S$.\\
From Eq. (\ref{schroedinger2}) we can obtain the following
continuity equation:
\begin{eqnarray}
\frac{\partial\,\rho}{\partial\,t}+{\bfm\nabla}\cdot{\bfm J}=0 \ ,
\end{eqnarray}
where the current ${\bfm J}$ now takes the standard expression of
the linear quantum mechanics given by Eq. (\ref{corcor}).

In conclusion, we have introduced a nonlinear and nonlocal
transformation which makes real the complex nonlinearity in Eq.
(\ref{schroedinger1}) and at the same time reduces the continuity
equation (\ref{continuity}) to the bilinear standard form. The
price that we pay is that Eq. (\ref{schroedinger2}) is not
generally canonical because the transformation (\ref{trasf1}) is
itself not canonical.

We briefly discuss the conditions under which the system,
described by Eq. (\ref{schroedinger2}), becomes a canonical one.
The canonicity of the system implies the existence of a nonlinear
potential $\widetilde U$ from which we can derive the nonlinearity
of Eq. (\ref{schroedinger2}).\\ We observe that the absence of the
imaginary part $\widetilde{\cal W}$ in the nonlinearity of Eq.
(\ref{schroedinger2}) requires that $\widetilde U$ depends only on
the field $\rho$ and its spatial derivatives. Consequently,
$\widetilde W$ is a functional of the field $\rho$ linked with
$\widetilde U(\rho)$ through the relation
\begin{eqnarray}
\widetilde{W}[\rho]=\frac{\delta}{\delta\,\rho}\,\int {\widetilde
U}[\rho]\,d^nx\,dt \ ,
\end{eqnarray}
which after performing the functional derivative assumes the form:
\begin{eqnarray}
\widetilde W[\rho]=\sum_{[k=0]}(-1)^k{\cal D}_{_{I_k}}\left[
\frac{\partial\,{\widetilde U}[\rho]}{\partial\,({\cal
D}_{_{I_k}}\rho)}\right] \ .\label{derfun1}
\end{eqnarray}
In section VI we will consider few particular systems where the
condition (\ref{derfun1}) is satisfied and their canonicity is
preserved.
\sect{Noncanonical Systems} In this section we consider the
transformation introduced previously and study its applicability
in the case of noncanonical systems.

For noncanonical systems, the evolution equation is given by Eq.
(\ref{sch1}), where $W[\rho,\,S]$ is now an arbitrary functional,
while $\cal{W}[\rho,\,S]$ assumes the form:
\begin{eqnarray}
{\cal W}[\rho,\,S]={\hbar\over2\,\rho}\,{\bfm\nabla}\cdot{\bfm
F}[\rho,\,S] \ ,\label{w}
\end{eqnarray}
enforced by the conservation of $N=\int\rho\,d^nx$, with ${\bfm
F}[\rho,\,S]$ an arbitrary functional, and the current is given
by:
\begin{eqnarray}
{\bfm j}=\frac{{\bfm\nabla}\,S}{m}\,\rho-{\bfm F}[\rho,\,S] \
.\label{current}
\end{eqnarray}

It is easy to verify that the transformation (\ref{trasf1})
eliminates the imaginary part of the nonlinearity in the motion
equation, which transforms again into Eqs.
(\ref{schroedinger2})-(\ref{real1}). The generator $\sigma$ of the
transformation $\psi\rightarrow\phi$ is related to $\bfm F$
through:
\begin{eqnarray}
{\bfm\nabla}\,\sigma[\rho,\,S]={m\over\rho}\,{\bfm F}[\rho,\,S] \
,\label{u1}
\end{eqnarray}
while the condition
\begin{eqnarray}
{\bfm\nabla}\times\frac{\bfm F}{\rho}=0 \ ,\label{condition}
\end{eqnarray}
permits the definition of the transformation in any $n>1$ spatial
dimension. This condition constrains only the form of $\cal W$,
differently to the canonical case, where the condition (\ref{rot})
constrains the form of the nonlinear potential $U$ and
consequently, both $W$ and $\cal W$ in the motion equation.

\sect{Examples} In this section we consider some equations already
known in literature in the framework of the approach developed
here. We show how the nonlinear transformations proposed to study
the various NLSEs, can be obtained in a unified way as particular
cases of the transformation given by Eq. (\ref{trasf1}).

\noindent {\bf i)} As first trivial example we consider the
canonical NLSE introduced in Ref. \cite{Bang1}:
\begin{eqnarray}
i\,\hbar\,\frac{\partial\,\psi}{\partial\,t}=-\frac{\hbar^2}{2\,m}\,\Delta\,\psi+\left(\beta\,\rho-
\frac{\bfm\alpha}{\hbar}\cdot{\bfm\nabla}\,S\right)\,\psi+\frac{i}{2}\,
\left(\bfm\alpha\cdot{\bfm\nabla}\log\rho\right)\,\psi \ ,
\end{eqnarray}
with $\beta$ and $\bfm \alpha$ real and constant arbitrary
parameters. This equation can be derived from a Lagrangian
containing the following nonlinear potential:
\begin{eqnarray}
U[\rho,\,S]=\frac{\beta}{2}\,\rho^2-\frac{\rho}{\hbar}\,{\bfm\alpha}\cdot{\bfm\nabla}\,S
\ .
\end{eqnarray}
The transformation with generator $\sigma$, given by:
\begin{eqnarray}
\sigma=-\frac{m}{\hbar}\,{\bfm\alpha}\cdot{\bfm x} \ ,
\end{eqnarray}
produces the new canonical evolution equation:
\begin{eqnarray}
i\,\hbar\,\frac{\partial\,\phi}{\partial\,t}=-\frac{\hbar^2}{2\,m}\,\Delta\,\phi
+\beta\,\rho\,\phi+\frac{m\,{\bfm\alpha}^2}{2\,\hbar^2}\,\phi \
,\label{eee1}
\end{eqnarray}
with associated nonlinear potential:
\begin{eqnarray}
{\widetilde U}[\rho]=\frac{\beta}{2}\,\rho^2+\frac{m\,{\bfm
\alpha}^2}{2\,\hbar}\,\rho \ .
\end{eqnarray}
Eq. (\ref{eee1}) can be reduced to the cubic NLSE by changing the
phase ${\cal S}\rightarrow{\cal
S}-3\,m\,{\bfm\alpha}^2\,t/2\,\hbar$.\\

\noindent {\bf ii)} Let us consider as second example of canonical
NLSE the Chen-Lee-Liu equation \cite{Chen}:
\begin{eqnarray}
i\,\hbar\,\frac{\partial\,\psi}{\partial\,t}=-\frac{\hbar^2}{2\,m}\,\frac{\partial^2\,\psi}{\partial\,x^2}
-{\alpha\over\hbar}\,\frac{\partial\,S}{\partial\,x}\,\rho\,\psi
+i\,\frac{\alpha}{2}\,\frac{\partial\,\rho}{\partial\,x}\,\psi \
,\label{exe1}
\end{eqnarray}
with $\alpha$ a real coupling constant. The nonlinear potential
associated to this equation is:
\begin{eqnarray}
U[\rho,\,S]=-\frac{\alpha}{2\,\hbar}\,\frac{\partial\,S}{\partial\,x}\,\rho^2
\ ,
\end{eqnarray}
while the transformation with generator:
\begin{eqnarray} \sigma=-\frac{\alpha\,m}{2\,\hbar}\int\rho\,dx
\ ,\label{sig1}
\end{eqnarray}
reduces Eq. (\ref{exe1}) to the following noncanonical NLSE:
\begin{eqnarray}
i\,\frac{\partial\,\phi}{\partial\,t}=-\frac{\partial^2\,\phi}{\partial\,x^2}-\frac{\alpha}{\hbar}\,\left(
\frac{\partial\,\cal
S}{\partial\,x}+{3\,\alpha\,m\over8\,\hbar}\,\rho\right)\,\rho\,\phi
\ .
\end{eqnarray}
We note that the transformation with generator $\sigma$ given by
Eq. (\ref{sig1}) is a particular case of the Kundu transformation
introduced in Ref. \cite{Kundu}.

\noindent {\bf iii)} As third example we consider the canonical
NLSE introduced in Ref.s \cite{Jackiw,Aglietti}:
\begin{eqnarray}
i\,\hbar\,\frac{\partial\,\psi}{\partial\,t}=-\frac{\hbar^2}{2\,m}\,\frac{\partial^2\,\psi}{\partial\,x^2}
-\frac{\lambda}{m}\,\left(
\frac{\lambda}{8}\,\rho+\frac{\partial\,S}{\partial\,x}\right)\,\rho\,\psi+i\,\frac{\hbar\,\lambda}{2\,m}\,
\frac{\partial\,\rho}{\partial\,x}\,\psi \ .\label{jackiw}
\end{eqnarray}
The associated potential is given by:
\begin{eqnarray}
U[\rho,\,S]=-\frac{3\,\lambda^2}{8\,m}\,
\rho^3-\frac{\lambda}{2\,m}\,\frac{\partial\,S}{\partial\,x}\,\rho^2
\ ,
\end{eqnarray}
while the generator $\sigma$ assumes the form:
\begin{eqnarray}
\sigma=-\frac{\lambda}{2}\int\rho\,dx \ .
\end{eqnarray}
The evolution equation for the field $\phi$ becomes:
\begin{eqnarray}
i\,\hbar\,\frac{\partial\,\phi}{\partial\,t}=-\frac{\hbar^2}{2\,m}\,\frac{\partial^2\,\phi}
{\partial\,x^2}-\frac{\lambda}{m}\,\frac{\partial\,{\cal S}
}{\partial\,x}\,\rho\,\phi \ .\label{exe3}
\end{eqnarray}

\noindent {\bf iv)} As fourth example we consider the canonical
NLSE recently introduced in Ref. \cite{Kaniadakis,Kaniadakis1}:
\begin{eqnarray}
i\,\hbar\,\frac{\partial\,\psi}{\partial\,t}=-\frac{\hbar^2}{2\,m}\,\frac{\partial^2\,\psi}{\partial\,x^2}
+\frac{\kappa}{m}\,\rho\,\left(\frac{\partial\,S}{\partial\,x}\right)^2\,\psi
-i\,\frac{\kappa\,\hbar}{2\,m\,\rho}\,\frac{\partial}{\partial\,x}
\left(\rho^2\,\frac{\partial\,S}{\partial\,x}\right)\,\psi \
.\label{exe4}
\end{eqnarray}
The nonlinear potential associated to Eq. (\ref{exe4}) is given
by:
\begin{eqnarray}
U[\rho,\,S]=\frac{\kappa}{2\,m}\,\left(\rho\,\frac{\partial\,S}{\partial\,x}\right)^2
\ .\label{ueip}
\end{eqnarray}
Although Eq. (\ref{exe4}) can be generalized in any spatial
dimension, it is easy to verify that the condition (\ref{rot}) is
not satisfied in this case, thus we can apply the transformation
only to the one-dimensional case. We perform the transformation
generated by:
\begin{eqnarray}
\sigma=\kappa\int\rho\,\frac{\partial\,S}{\partial\,x}\,dx \ ,
\end{eqnarray}
and Eq. (\ref{exe4}) transforms into:
\begin{eqnarray}
i\,\hbar\,\frac{\partial\,\phi}{\partial\,t}=-\frac{\hbar^2}{2\,m}\,\frac{\partial^2\,\phi}{\partial\,x^2}+
\frac{\kappa}{m}\,\frac{\rho}{1+\kappa\,\rho}\,\left(\frac{\partial\,{\cal
S}}{\partial\,x}\right)^2\,\phi
-\kappa\,\frac{\hbar^2}{4\,m}\,\rho\,\frac{\partial^2\,\log\rho}{\partial\,x^2}\,\phi
\ .\label{exe41}
\end{eqnarray}

\noindent {\bf v)} As last example of canonical system we consider
the sub-class of DG equations given by \cite{Doebner1}:
\begin{eqnarray}
i\,\hbar\,\frac{\partial\,\psi}{\partial\,t}=-\frac{\hbar^2}{2\,m}\,\Delta\,\psi
+\left\{\alpha\,\Delta\,S
-2\,\beta\,\frac{\hbar^2}{m}\,\left[\frac{\Delta\rho}{\rho}-{1\over2}\,\left(\frac{{\bfm\nabla}\,\rho}{\rho}
\right)^2\right]\right\}\,\psi+i\,\alpha\,\frac{\hbar}{2}\,\frac{\Delta\rho}{\rho}\,\psi
\ ,\label{dg}
\end{eqnarray}
with associated a nonlinear potential:
\begin{eqnarray}
U[\rho,\,S]=\alpha\,\rho\,\Delta\,S+\beta\,\frac{\hbar^2}{m}
\frac{({\bfm\nabla}\,\rho)^2}{\rho} \ .\label{dgpotential}
\end{eqnarray}
The generator $\sigma$ now is:
\begin{eqnarray}
\sigma=-m\,\alpha\,\log\rho \ ,\label{ss}
\end{eqnarray}
while the evolution equation for the field $\phi$ becomes:
\begin{eqnarray}
i\,\hbar\,\frac{\partial\,\phi}{\partial\,t}=-\frac{\hbar^2}{2\,m}\,\Delta\,\phi+\gamma\,
\left[\frac{\Delta\rho}{\rho}-{1\over2}\,\left(\frac{{\bfm\nabla}\,\rho}{\rho}
\right)^2\right]\,\phi \ ,
\end{eqnarray}
with $\gamma=m\,\alpha^2-2\,\beta\,\hbar^2/m$. This equation is
again canonical with nonlinear potential:
\begin{eqnarray}
{\widetilde
U}[\rho]=-\frac{\gamma}{2}\,\frac{({\bfm\nabla}\,\rho)^2}{\rho} \
,
\end{eqnarray}
and can be linearized performing the rescaling ${\cal
S}\rightarrow{\cal S}\,\sqrt{2\,m\,\gamma/\hbar^2-1}$, as noted in
Ref.\cite{Guerra}.

\noindent {\bf vi)} The most general class of DG equations is non
canonical and takes the form:
\begin{eqnarray}
i\,\hbar\,\frac{\partial\,\psi}{\partial\,t}=-\frac{\hbar^2}{2\,m}\,\Delta\,\psi
+\hbar\,D^\prime\,\sum_{i=1}^5 c_i\,R_i[\rho,\,S]\,\psi+i\,
\frac{\hbar}{2}\,D\,R_2[\rho,\,S]\,\psi \ ,\label{dg1}
\end{eqnarray}
where $R_1={\bfm\nabla}\cdot{\bfm
j}/\rho,\,R_2=\Delta\rho/\rho,\,R_3=({\bfm j}/\rho)^2,\,R_4={\bfm
j}\cdot{\bfm\nabla}\rho/\rho^2,\,R_5=({\bfm\nabla}\rho/\rho)^2$.
Note that Eq. (\ref{dg}) is obtained when $D=\alpha,\,c
_1=-c_4=m\,\alpha/\hbar\,D^\prime,\,c_3=0$ and
$c_2=-2\,c_5=-2\,\beta\,\hbar/m\,D^\prime$. The same generator
$\sigma$ given by Eq. (\ref{ss}) defines the transformation
$\psi\rightarrow\phi$ reducing the evolution equation to:
\begin{eqnarray}
i\,\hbar\,\frac{\partial\,\phi}{\partial\,t}=-\frac{\hbar^2}{2\,m}\,\Delta\,\phi+
\sum_{i=1}^5 \tilde{c}_i\,R_i[\rho,\,{\cal S}]\,\phi \ ,
\end{eqnarray}
where now the coefficients are given by
$\tilde{c}_1=\hbar\,D^\prime\,c_1-m\,D,\,\tilde{c}_2=\hbar\,D^\prime\,c_2,\,
\tilde{c}_3=\hbar\,D^\prime\,c_3,\,\tilde{c}_4=\hbar\,D^\prime\,c_4+m\,D,\,
\tilde{c}_5=\hbar\,D^\prime+m\,D^2/2$. Note that in Ref.
\cite{Doebner2} a nonlinear transformation was introduced with
generator:
\begin{eqnarray}
\sigma_{_{\rm DG}}=\frac{\gamma(t)}{2}\,\log{\rho}+
\frac{1}{\hbar}\,[\lambda(t)-1]\,S+\theta(t,\,{\bfm x}) \
,\label{dbt}
\end{eqnarray}
which produces a group of transformations mapping the DG equation
into itself. We observe that, after posing in Eq. (\ref{dbt})
$\theta(t,{\bfm x})=0,\,\lambda(t)=1$ and
$\gamma(t)=2\,m\,\beta/\hbar$, we obtain the generator $\sigma$
given by Eq. (\ref{ss}).

\noindent {\bf vii)} As a second example of noncanonical system we
consider the equation:
\begin{eqnarray}
i\,\hbar\,\frac{\partial\,\psi}{\partial\,t}=-\frac{\hbar^2}{2\,m}\,\frac{\partial^2\,\psi}{\partial\,x^2}
+i\,\alpha(\psi^\ast\,\frac{\partial\,\psi}{\partial\,x}+q\,\psi\,\frac{\partial\,\psi^\ast}{\partial\,x})\,\psi
\ ,\label{exe11}
\end{eqnarray}
with $q$ a real parameter. We note that for $q=1/2$, Eq.
(\ref{exe11}) reduces to the Kaup-Newell equation \cite{Kaup},
while for $q=0$ we obtain the Chen-Lee-Liu equation (\ref{exe1}).
Finally for $q=-1$ the nonlinearity in Eq. (\ref{exe11}) becomes
purely real and the equation coincides with Eq. (\ref{exe3})
obtained previously. The generator:
\begin{eqnarray}
\sigma=-\frac{m\,\alpha}{2\,\hbar}\,(q+1)\,\int\rho\,dx \
,\label{gen1}
\end{eqnarray}
defines a transformation which reduces Eq. (\ref{exe11}) to:
\begin{eqnarray}
i\,\hbar\,\frac{\partial\,\phi}{\partial\,t}=-\frac{\hbar^2}{2\,m}\,\frac{\partial^2\,\phi}{\partial\,x^2}
-\frac{\alpha}{\hbar}\left[(1-q)\,\frac{\partial\,{\cal
S}}{\partial\,x}
+{1\over8}\,m\,\frac{\alpha}{\hbar}\,(3-2\,q-5\,q^2)\,\rho\right]\,\rho\,\phi
\ .\label{exe12}
\end{eqnarray}
In the particular case $q=1$ Eq. (\ref{exe12}) becomes canonical
with nonlinear potential:
\begin{eqnarray}
{\widetilde U}[\rho]=\frac{m\,\alpha^2}{6\,\hbar^2}\,\rho^3 \ .
\end{eqnarray}

\noindent {\bf viii)} As last example we consider the Eckaus
equation \cite{Calogero1} which is a noncanonical NLSE:
\begin{eqnarray}
i\,\hbar\,\frac{\partial\,\psi}{\partial\,t}=-\frac{\hbar^2}{2\,m}\,\frac{\partial^2\,\psi}{\partial\,x^2}
+i\,\alpha\,\frac{\partial\,\rho}{\partial\,x}\,\psi+\beta\,\rho^2\,\psi
\ .\label{exe2}
\end{eqnarray}
The generator:
\begin{eqnarray}
\sigma=-\frac{m\,\alpha}{\hbar}\,\int\rho\,dx \ ,\label{sss}
\end{eqnarray}
defines the transformation reducing Eq. (\ref{exe2}) to the
well-known quintic NLSE:
\begin{eqnarray}
i\,\hbar\,\frac{\partial\,\phi}{\partial\,t}=-\frac{\hbar^2}{2\,m}\,\frac{\partial^2\,\phi}{\partial\,x^2}
+\left(\frac{m\,\alpha^2}{2\,\hbar^2}+\beta\right)\,\rho^2\,\phi \
.\label{exe212}
\end{eqnarray}
Eq. (\ref{exe212}) is a canonical one with nonlinear potential:
\begin{eqnarray}
{\widetilde
U}[\rho]={1\over3}\,\left(\frac{m\,\alpha^2}{2\,\hbar^2}+\beta\right)\,\rho^3
\ ,
\end{eqnarray}
In the particular case where $\beta=-m\,\alpha^2/2\,\hbar^2$ the
transformation with generator given by Eq. (\ref{sss}) linearizes
Eq. (\ref{exe2}).
\sect{Conclusion} In this paper we have consider a class of
canonical NLSEs containing complex nonlinearities and describing
$U(1)$-invariant systems. For these systems we study the
symmetries and the conserved quantities associated to
roto-translations and Galilei invariance.\\ Subsequently, we
introduce a Cole-Hopf like transformation $\psi\rightarrow{\cal
U}\,\psi$, which preserves the $U(1)$-invariance of the system and
reduces the complex nonlinearity into a real one so that the
continuity equation assumes the standard bilinear form. This
transformation generally does not conserve the canonicity of the
system. Extension to noncanonical equations is also studied.

The general Cole-Hopf-like transformation introduced here, allows
us to deal in a unifying scheme several NLSEs already known in
literature, obtaining, in this way, the transformations
introduced by various authors.
\app
\section{} In this Appendix we recover, by using the Noether theorem \cite{Noether}, the
continuity equation associated to a given symmetry.

Let us consider the action:
\begin{eqnarray}
{\cal A}=\int{\cal L}\,d^nx\,dt \ ,\label{action1}
\end{eqnarray}
with Lagrangian:
\begin{eqnarray}
{\cal L}={\cal L}_{_{\rm L}}+{\cal L}_{_{\rm NL}} \ ,
\end{eqnarray}
where
\begin{eqnarray}
{\cal L}_{_{\rm L}}
=i\,\frac{\hbar}{2}\left(\psi^\ast\,\frac{\partial\,\psi}{\partial\,t}
-\psi\,\frac{\partial\,\psi^\ast}{\partial\,t}\right)-\frac{\hbar^2}{2\,m}\,|{\bfm\nabla}\,\psi|^2
\ ,
\end{eqnarray}
the standard Lagrangian density of the linear Schr\"odinger theory
while
\begin{eqnarray}
{\cal L}_{_{\rm NL}}=-U[\rho,\,S] \ ,\label{unl}
\end{eqnarray}
is a real scalar functional depending on the hydrodynamic fields
$\rho,\,S$ and their spatial derivatives. The evolution equation
for the field $\psi$ is given by:
\begin{eqnarray}
\frac{\partial\,{\cal A}}{\partial\,\psi^\ast}=0 \ .\label{act}
\end{eqnarray}
Taking the functional derivatives Eq. (\ref{act}) becomes:
\begin{eqnarray}
\nonumber &&\frac{\partial\,{\cal L}_{_{\rm
L}}}{\partial\,\psi^\ast}-\frac{\partial}{\partial\,t}\,
\frac{\partial\,{\cal L}_{_{\rm
L}}}{\partial\,(\partial_t\psi^\ast)}
-\frac{\partial}{\partial\,x_i}\,\frac{\partial\,{\cal L}_{_{\rm
L}}}{\partial\,(\partial_i\psi^\ast)}+ \sum_{[k=0]}\,(-1)^k\,{\cal
D}_{_{I_k}}\left[\frac{\partial\,{\cal L}_{_{\rm
NL}}}{\partial\,({\cal D}_{_{I_k}}\rho)}\right]\,\psi\\
&&+i\,\frac{\hbar}{2\,\rho}\, \sum_{[k=0]}\,(-1)^k\,{\cal
D}_{_{I_k}}\left[\frac{\partial\,{\cal L}_{_{\rm
NL}}}{\partial\,({\cal D}_{_{I_k}}S)}\right]\,\psi=0 \
.\label{aa5}
\end{eqnarray}

We compute the variation $\delta_\epsilon\,{\cal A}$ generated by
a one-parameter transformation group. For simplicity, we assume
that the symmetry group acts only on the internal degrees of
freedom of the system. The contributions to
$\delta_\epsilon\,{\cal A}$, given by the variation of the volume
element $d^nx\,dt$, when the symmetry involves the space-time
variables, are well known and can be added successively. Thus, we
have:
\begin{eqnarray}
\nonumber \delta_\epsilon\,{\cal
A}\,\,&&=\int\left[\frac{\delta\,{\cal L}_{_{\rm
L}}}{\delta\,\psi}\,\delta_\epsilon\,\psi+\frac{\delta\,{\cal
L}_{_{\rm
L}}}{\delta\,\psi^\ast}\,\delta_\epsilon\,\psi^\ast+\frac{\delta\,{\cal
L}_{_{\rm
NL}}}{\delta\,\rho}\,\delta_\epsilon\,\rho+\frac{\delta\,{\cal
L}_{_{\rm NL}}}{\delta\,S}\,\delta_\epsilon\,S\right]\,d^nx\,dt\\
\nonumber &&=\int\Bigg\{\frac{\partial\,{\cal L}_{_{\rm
L}}}{\partial\,\psi}\,\delta_\epsilon\,\psi+\frac{\partial\,{\cal
L}_{_{\rm
L}}}{\partial\,(\partial_t\psi)}\,\delta_\epsilon\,(\partial_t\psi)+\frac{\partial\,{\cal
L}_{_{\rm
L}}}{\partial\,(\partial_i\psi)}\,\delta_\epsilon\,(\partial_i\psi)\\
\nonumber &&+\,\,\frac{\partial\,{\cal L}_{_{\rm
L}}}{\partial\,\psi^\ast}\,\delta_\epsilon\,\psi^\ast+\frac{\partial\,{\cal
L}_{_{\rm
L}}}{\partial\,(\partial_t\psi^\ast)}\,\delta_\epsilon\,(\partial_t\psi^\ast)+\frac{\partial\,{\cal
L}_{_{\rm
L}}}{\partial\,(\partial_i\psi^\ast)}\,\delta_\epsilon\,(\partial_i\psi^\ast)\\
&&+\sum_{[k=0]}\left[\frac{\partial\,{\cal L}_{_{\rm
NL}}}{\partial\,({\cal D}_{_{I_k}}\rho)}\,\delta_\epsilon\,({\cal
D}_{_{I_k}}\rho)+\frac{\partial\,{\cal L}_{_{\rm
NL}}}{\partial\,({\cal D}_{_{I_k}}S)}\,\delta_\epsilon\,({\cal
D}_{_{I_k}}S)\right] \Bigg\} \,d^nx\,dt \ ,\label{a1}
\end{eqnarray}
with ${\cal D}_{_{I_k}}\equiv\partial^k/(\partial
x_{_1}^{i_1}\cdots x_{_n}^{i_n})$ and the Einstein convention for
the repeated indices is assumed. In Eq. (\ref{a1}) we have posed
$\sum_{[k=0]}\equiv\sum_{k=0}^\infty\,\sum_{I_k}$, where the
second sum is performed on the multi-index
$I_k\equiv(i_1,\,i_2,\,\cdots,\,i_n)$ with $0\leq i_p\leq k$,
$\sum i_p=k$. If we use the identity:
\begin{eqnarray}
\frac{\partial\,{\cal
L}}{\partial\,(\partial_a\phi)}\,\delta_\epsilon\,(\partial_a\phi)=
\frac{\partial}{\partial\,a}\,\left[\frac{\partial\,{\cal
L}}{\partial\,(\partial_a\phi)}\,\delta_\epsilon\,\phi\right]-\frac{\partial}{\partial\,a}\,
\left[\frac{\partial\,{\cal
L}}{\partial\,(\partial_a\phi)}\right]\,\delta_\epsilon\,\phi \
,\label{iden}
\end{eqnarray}
with $a\equiv t,\,i$, Eq. (\ref{a1}) becomes:
\begin{eqnarray}
\nonumber \delta_\epsilon\,{\cal
A}=&&\int\Bigg\{\frac{\partial\,{\cal L}_{_{\rm
L}}}{\partial\,\psi}\,\delta_\epsilon\,\psi+\frac{\partial}{\partial\,t}\,\left[\frac{\partial\,{\cal
L}_{_{\rm
L}}}{\partial\,(\partial_t\psi)}\,\delta_\epsilon\,\psi\right]-\frac{\partial}{\partial\,t}\,
\left[\frac{\partial\,{\cal L}_{_{\rm
L}}}{\partial\,(\partial_t\psi)}\right]\,\delta_\epsilon\,\psi+\frac{\partial}{\partial\,x_i}
\,\left(\frac{\partial\,{\cal L}_{_{\rm
L}}}{\partial\,(\partial_i\psi)}\,\delta_\epsilon\,\psi\right)\\
\nonumber &&-\,\,\frac{\partial}{\partial\,x_i}\,
\left[\frac{\partial\,{\cal L}_{_{\rm
L}}}{\partial\,(\partial_i\psi)}\right]\,\delta_\epsilon\,\psi+\frac{\partial\,{\cal
L}_{_{\rm L}}}{\partial\,\psi^\ast}\,\delta_\epsilon\,\psi^\ast +
\frac{\partial}{\partial\,t}\,\left[\frac{\partial\,{\cal
L}_{_{\rm
L}}}{\partial\,(\partial_t\psi^\ast)}\,\delta_\epsilon\,\psi^\ast\right]-\frac{\partial}{\partial\,t}\,
\left[\frac{\partial\,{\cal L}_{_{\rm
L}}}{\partial\,(\partial_t\psi^\ast)}\right]\,\delta_\epsilon\,\psi^\ast\\
\nonumber &&+\,\,\frac{\partial}{\partial\,x_i}
\,\left[\frac{\partial\,{\cal L}_{_{\rm
L}}}{\partial\,(\partial_i\psi^\ast)}\,\delta_\epsilon\,\psi^\ast\right]
-\frac{\partial}{\partial\,x_i}\, \left[\frac{\partial\,{\cal
L}_{_{\rm
L}}}{\partial\,(\partial_i\psi^\ast)}\right]\,\delta_\epsilon\,\psi^\ast
\\&&+\sum_{[k=0]}\left[\frac{\partial\,{\cal L}_{_{\rm
NL}}}{\partial\,({\cal D}_{_{I_k}}\rho)}\,\delta_\epsilon\,({\cal
D}_{_{I_k}}\rho)+\frac{\partial\,{\cal L}_{_{\rm
NL}}}{\partial\,({\cal D}_{_{I_k}}S)}\,\delta_\epsilon\,({\cal
D}_{_{I_k}}S)\right] \Bigg\} \,d^nx\,dt \ .\label{a2}
\end{eqnarray}

For a fixed value of the index $k$ and multi-index $I_k$, using
$k$ times Eq. (\ref{iden}), we have:
\begin{eqnarray}
\frac{\partial\,{\cal L}_{_{\rm NL}}}{\partial\,({\cal
D}_{_{I_k}}\rho)}\,\delta_\epsilon\,({\cal
D}_{_{I_k}}\rho)=\sum_{[p=0]}^k\,(-1)^p\,A_{_{I_q}}^{^{I_k}}\,{\cal
D}_{_{I_q}}\left[{\cal D}_{_{I_p}} \left(\frac{\partial\,{\cal
L}_{_{\rm NL}}}{\partial\,({\cal
D}_{_{I_k}}\rho)}\right)\,\delta_\epsilon\,\rho\right] \
,\label{a4}
\end{eqnarray}
where the coefficient
$A_{_{I_q}}^{^{I_k}}=\prod_{r=1}^ni_r!/(l_r!\,m_r!)$,
$\sum_{[p=0]}^k\equiv\sum_{p=0}^k\,\sum_{_{I_p}}$ and the
multi-index $I_k=(i_1,\cdots,i_n),\,\,I_p=(l_1,\cdots,l_n)$ and
$I_q=(m_1,\cdots,m_n)$ are related by $i_r=l_r+m_r$.

Using Eq. (\ref{a4}), Eq. (\ref{a2}) transforms to:
\begin{eqnarray}
\nonumber \delta_\epsilon\,{\cal
A}&&\,\,=\int\Bigg\{\frac{\partial\,{\cal L}_{_{\rm
L}}}{\partial\,\psi}\,\delta_\epsilon\,\psi+\frac{\partial}{\partial\,t}\,\left[\frac{\partial\,{\cal
L}_{_{\rm
L}}}{\partial\,(\partial_t\psi)}\,\delta_\epsilon\,\psi\right]-\frac{\partial}{\partial\,t}\,
\left[\frac{\partial\,{\cal L}_{_{\rm
L}}}{\partial\,(\partial_t\psi)}\right]\,\delta_\epsilon\,\psi+\frac{\partial}{\partial\,x_i}
\,\left[\frac{\partial\,{\cal L}_{_{\rm
L}}}{\partial\,(\partial_i\psi)}\,\delta_\epsilon\,\psi\right]\\
\nonumber &&-\,\,\frac{\partial}{\partial\,x_i}\,
\left[\frac{\partial\,{\cal L}_{_{\rm
L}}}{\partial\,(\partial_i\psi)}\right]\,\delta_\epsilon\,\psi+\frac{\partial\,{\cal
L}_{_{\rm L}}}{\partial\,\psi^\ast}\,\delta_\epsilon\,\psi^\ast +
\frac{\partial}{\partial\,t}\,\left[\frac{\partial\,{\cal
L}_{_{\rm
L}}}{\partial\,(\partial_t\psi^\ast)}\,\delta_\epsilon\,\psi^\ast\right]-\frac{\partial}{\partial\,t}\,
\left[\frac{\partial\,{\cal L}_{_{\rm
L}}}{\partial\,(\partial_t\psi^\ast)}\right]\,\delta_\epsilon\,\psi^\ast
\\ \nonumber &&+\,\,\frac{\partial}{\partial\,x_i}
\,\left[\frac{\partial\,{\cal L}_{_{\rm
L}}}{\partial\,(\partial_i\psi^\ast)}\,\delta_\epsilon\,\psi^\ast\right]
-\frac{\partial}{\partial\,x_i}\, \left[\frac{\partial\,{\cal
L}_{_{\rm
L}}}{\partial\,(\partial_i\psi^\ast)}\right]\,\delta_\epsilon\,\psi^\ast\\
&&+\sum_{[k=0]}\,\sum_{[p=0]}^k\,(-1)^p\,A_{_{I_q}}^{^{I_k}}\,{\cal
D}_{_{I_q}}\Bigg\{{\cal D}_{_{I_p}} \left[\frac{\partial\,{\cal
L}_{_{\rm NL}}}{\partial\,({\cal
D}_{_{I_k}}\rho)}\right]\,\delta_\epsilon\,\rho +{\cal D}_{_{I_p}}
\left[\frac{\partial\,{\cal L}_{_{\rm NL}}}{\partial\,({\cal
D}_{_{I_k}}S)}\right]\,\delta_\epsilon\,S\Bigg\} \Bigg\}
\,d^nx\,dt \ .\label{a5}
\end{eqnarray}
After inserting in Eq. (\ref{a5}) the expression of
$\partial\,{\cal L}_{_{\rm NL}}/\partial\,\psi$ and
$\partial\,{\cal L}_{_{\rm NL}}/\partial\,\psi^\ast$ obtained from
Eq. (\ref{aa5}) and its conjugate, we finally obtain:
\begin{eqnarray}
\nonumber \delta_\epsilon\,{\cal
A}&&\,\,=\int\Bigg\{\frac{\partial}{\partial\,t}\,\left[\frac{\partial\,{\cal
L}_{_{\rm
L}}}{\partial\,(\partial_t\psi)}\,\delta_\epsilon\,\psi+\frac{\partial\,{\cal
L}_{_{\rm
L}}}{\partial\,(\partial_t\psi^\ast)}\,\delta_\epsilon\,\psi^\ast\right]+\frac{\partial}{\partial\,x_i}
\,\left[\frac{\partial\,{\cal L}_{_{\rm
L}}}{\partial\,(\partial_i\psi)}\,\delta_\epsilon\,\psi+\frac{\partial\,{\cal
L}_{_{\rm
L}}}{\partial\,(\partial_i\psi^\ast)}\,\delta_\epsilon\,\psi^\ast\right]\\
&&+\sum_{[k=1]}\,\sum_{[p=0]}^{k-1}\,(-1)^p\,A_{_{I_q}}^{^{I_k}}\,{\cal
D}_{_{I_q}}\Bigg\{{\cal D}_{_{I_p}} \left[\frac{\partial\,{\cal
L}_{_{\rm NL}}}{\partial\,({\cal
D}_{_{I_k}}\rho)}\right]\,\delta_\epsilon\,\rho +{\cal D}_{_{I_p}}
\left[\frac{\partial\,{\cal L}_{_{\rm NL}}}{\partial\,({\cal
D}_{_{I_k}}S)}\right]\,\delta_\epsilon\,S\Bigg\} \Bigg\}
\,d^nx\,dt \ .\label{a6}
\end{eqnarray}
In presence of a symmetry the variation of the action must vanish
and thus, from Eq. (\ref{a6}), after rearranging the terms, it
follows the continuity equation:
\begin{eqnarray}
\frac{\partial\,{\cal Q}}{\partial\,t}+{\bfm\nabla}\cdot{\bfm
{\cal F}}=0 \ ,\label{a7}
\end{eqnarray}
with charge $\cal Q$:
\begin{eqnarray}
{\cal Q}=\frac{\partial\,{\cal L}_{_{\rm
L}}}{\partial\,(\partial_t\psi)}\,\delta_\epsilon\,\psi+\frac{\partial\,{\cal
L}_{_{\rm
L}}}{\partial\,(\partial_t\psi^\ast)}\,\delta_\epsilon\,\psi^\ast
\ , \label{charge}
\end{eqnarray}
and the flux $\bfm{\cal F}$:
\begin{eqnarray}
\nonumber {\cal F}_j=&&\frac{\partial\,{\cal L}_{_{\rm
L}}}{\partial\,(\partial_j\psi)}\,\delta_\epsilon\,\psi+\frac{\partial\,{\cal
L}_{_{\rm
L}}}{\partial\,(\partial_j\psi^\ast)}\,\delta_\epsilon\,\psi^\ast\\
+&&\sum_{[k=0]}\,\sum_{[p=0]}^k\,(-1)^p\,B_{_{j,I_q}}^{^{I_k}}
\,{\cal D}_{_{I_q}}\Bigg\{{\cal D}_{_{I_p}}
\left[\frac{\partial\,{\cal L}_{_{\rm NL}}}{\partial\,({\cal
D}_{_{j,I_k}}\rho)}\right]\,\delta_\epsilon\,\rho +{\cal
D}_{_{I_p}} \left[\frac{\partial\,{\cal L}_{_{\rm
NL}}}{\partial\,({\cal
D}_{_{j,I_k}}S)}\right]\,\delta_\epsilon\,S\Bigg\} \ ,\label{flux}
\end{eqnarray}
where ${\cal D}_{_{j,I_k}}a\equiv\partial_j\,{\cal D}_{_{I_k}}a$
with $a\equiv\rho,\,S$, the coefficients
$B_{_{j,I_q}}^{^{I_k}}=(i_j+1)\,A_{_{I_q}}^{^{I_k}}/(m_j+1)\,
f_j^{_{I_q}}$ and $f_j^{_{I_q}}=n-\sum_{r\not=j}^n\delta_{0,m_r}$.
Recall that from the continuity equation the current is defined
modulo the curl of an arbitrary function. This fact was taken into
account in the expression of the current (\ref{flux}).

In the following we discuss two important cases. In the first, we
suppose the system is $U(1)$-invariant. Using the transformation
$\psi\rightarrow\psi\,\exp(i\,\epsilon)$ where $\epsilon$ is the
infinitesimal generator, we have:
\begin{eqnarray}
\begin{array}{ll}
\delta_\epsilon\,\psi=i\,\epsilon\,\psi \ ,
&\hspace{10mm}\delta_\epsilon\,\psi^\ast=-i\,\epsilon\,\psi^\ast \
,\\ \delta_\epsilon\,\rho=0 \
,&\hspace{10mm}\delta_\epsilon\,S=\hbar\,\epsilon \ .
\end{array}
\end{eqnarray}
From Eqs. (\ref{charge}) and (\ref{flux}) we obtain the expression
of the conserved density ${\cal Q}=\rho$ and the related current
${\cal F}_i=j_i$:
\begin{eqnarray}
j_i=\frac{\partial_i\,S}{m}\,\rho+
\sum_{[k=0]}\,\frac{(-1)^k}{f_i^{_{I_{k+1}}}}\,{\cal
D}_{_{I_k}}\left[\frac{\partial\,U[\rho,\,S]}{\partial\,({\cal
D}_{_{i,I_k}}S)}\right] \ .\label{currr}
\end{eqnarray}
It is trivial to note that Eq. (\ref{a7}) is the continuity
equation for the field $\psi$, where the current (\ref{currr})
assumes a nonstandard expression, due to the presence of the
imaginary part of the nonlinearity in the evolution equation. By
taking into account the definition (\ref{derfun}) of the
functional derivative, Eq. (\ref{currr}) can be written also in
the form:
\begin{eqnarray}
j_i=\frac{\partial_i\,S}{m}\,\rho+\frac{\delta}{\delta\,(\partial_i
S})\,\int U[\rho,\,S]\,d^nx\,dt \ ,\label{currr1}
\end{eqnarray}
modulo a curl of an arbitrary function. Note that Eq.
(\ref{currr1}) can be obtained starting directly from the Eq.
(\ref{schroedinger1}) after making the hypothesis that
$U[\rho,\,S]$ depends on the field $S$ only through its spatial
derivatives as required from the U(1) symmetry.

In the second case, we discuss the energy-momentum tensor related
to the space-time translations. Posing $x_\mu\rightarrow
x_\mu+\epsilon_\mu$ we have:
\begin{eqnarray}
\begin{array}{ll}
\delta_\epsilon\,\psi=\epsilon_\mu\,\partial_\mu\psi \ ,
&\hspace{10mm}\delta_\epsilon\,\psi^\ast=\epsilon_\mu\,\partial_\mu\psi^\ast
\ ,\\ \delta_\epsilon\,\rho=\epsilon_\mu\,\partial_\mu\rho \
,&\hspace{10mm}\delta_\epsilon\,S=\epsilon_\mu\partial_\mu\,S \ ,
\end{array}
\end{eqnarray}
with $\mu=0,\cdots,3$ and $\partial_0\equiv\partial_t$. From Eqs.
(\ref{charge}) and (\ref{flux}) we obtain:
\begin{eqnarray}
T_{00}=&& \frac{\hbar^2}{2\,m}\,|{\bfm\nabla}\,\psi|^2+U[\rho,\,S]
\ ,\label{t1}\\
T_{0j}=&&i\,\frac{\hbar}{2}\,(\psi^\ast\,\partial_j\,\psi
-\psi\,\partial_j\,\psi^\ast) \ ,\label{t2}\\ \nonumber
T_{i0}=&&-\frac{\hbar^2}{2\,m}\left(\partial_i\psi^\ast\,\partial_t\psi-
\partial_i\psi\,\partial_t\psi^\ast\right)
\\
&&+\sum_{[k=0]}\sum_{[p=0]}^k(-1)^pB_{_{j,I_q}}^{^{I_k}}{\cal
D}_{_{I_q}}\left\{{\cal
D}_{_{I_p}}\left[\frac{\partial\,U[\rho,\,S]}{\partial\left({\cal
D}_{_{i,I_k}}\rho\right)}\right]\,\partial_t\rho+{\cal
D}_{_{I_p}}\left[\frac{\partial\,U[\rho,\,S]}{\partial\,\left({\cal
D}_{_{i,I_k}}S\right)}\right]\,\partial_tS\right\}
 \ ,\label{t3}\\ \nonumber
T_{ij}=&&-\frac{\hbar^2}{2\,m}\left(\partial_i\psi^\ast\,\partial_j\psi+
\partial_j\psi^\ast\,\partial_i\psi\right)+\delta_{ij}{\cal
L}\\&&-\sum_{[k=0]} \sum_{[p=0]}^k(-1)^pB_{_{j,I_q}}^{^{I_k}}{\cal
D}_{_{I_q}}\left\{{\cal
D}_{_{I_p}}\left[\frac{\partial\,U[\rho,\,S]}{\partial\,\left({\cal
D}_{_{i,I_k}}\rho\right)}\right]\partial_j\rho+{\cal
D}_{_{I_p}}\left[\frac{\partial\,U[\rho,\,S]}{\partial\,\left({\cal
D}_{_{i,I_k}}S\right)}\right]\partial_jS\right\} \ .\label{t4}
\end{eqnarray}
In Eqs. (\ref{t1}) and (\ref{t4}) we have taken into account the
contribution due to the volume element. Note that the potential
$U[\rho,\,S]$ does not modify the expression of the momentum
density $T_{0j}$ which assumes the same form as in the linear
theory. In contrast, $U[\rho,\,S]$ changes the expression of the
energy density $T_{00}$, and more strongly the expression of the
flux densities $T_{i\mu}$.
\sect{} {\bf Theorem:} \it If $U[\rho,\,S]$ and ${\bfm
F}[\rho,\,S]$ are two smooth functionals depending on the fields
$\rho,\,S$ and their spatial derivatives and satisfy the relation:
\begin{eqnarray}
\frac{\partial}{\partial\,S}\,U[\rho,\,S]={\bfm\nabla}\cdot{\bfm
F}[\rho,\,S] \ ,\label{pro}
\end{eqnarray}
the functional $U[\rho,\,S]$ takes the form:
\begin{eqnarray}
U[\rho,\,S]=\overline U[\rho,\,S]+{\bfm\nabla}\cdot{\bfm
G}[\rho,\,S] \ ,\label{sol}
\end{eqnarray}
where $\overline U[\rho,\,S]$ depends on $S$ only through its
derivatives: $\partial\,\overline U/\partial\,S=0$. \rm

{\bf Proof:} Deriving Eq. (\ref{sol}) with respect to $S$ we
obtain:
\begin{eqnarray}
\nonumber
\frac{\partial\,U}{\partial\,S}=&&\frac{\partial\,\overline
U}{\partial\,S}+\frac{\partial}{\partial\,S}\,{\bfm\nabla}\,\cdot{\bfm
G}=
\frac{\partial}{\partial\,S}\,\sum_{[k=0]}\left[\partial_i({\cal
D}_{_{I_k}}\rho) \frac{\partial\,G_i}{\partial\,({\cal
D}_{_{I_k}}\rho)} +\partial_i({\cal D}_{_{I_k}}S)\frac{\partial
\,G_i}{\partial\,({\cal D}_{_{I_k}}S)}\right]\\
=&&\sum_{[k=0]}\left[\partial_i\,({\cal
D}_{_{I_k}}\rho)\frac{\partial}{\partial\,({\cal
D}_{_{I_k}}\rho)}+\partial_i({\cal
D}_{_{I_k}}S)\frac{\partial}{\partial\,({\cal
D}_{_{I_k}}S)}\right]\,\frac{\partial\,G_i}{\partial\,S}=
{\bfm\nabla}\cdot\frac{\partial\,{\bfm G}}{\partial\,S} \
,\label{prova1}
\end{eqnarray}
which coincides with Eq. (\ref{pro}) for ${\bfm F}=\partial\,{\bfm
G}/\partial\,S$.\\ Alternatively, expanding the r.h.s. of Eq.
(\ref{pro}):
\begin{eqnarray}
\frac{\partial\,U}{\partial\,S}=\sum_{[k=0]}\left[\partial_i({\cal
D}_{_{I_k}}\rho) \frac{\partial\,F_i}{\partial\,({\cal
D}_{_{I_k}}\rho)}+\partial_i({\cal
D}_{_{I_k}}S)\frac{\partial\,F_i}{\partial\,({\cal
D}_{_{I_k}}S)}\right] \ ,
\end{eqnarray}
and integrating on the field $S$, after taking into account that
$\rho,\,S,\,{\cal D}_{_{I_k}}\rho$ and ${\cal D}_{_{I_k}}S$ are
independent quantities, we have:
\begin{eqnarray}
\nonumber U=&&\sum_{[k=0]}\int\left[\partial_i({\cal
D}_{_{I_k}}\rho)\frac{\partial\,F_i}{\partial\,({\cal
D}_{_{I_k}}\rho)}+\partial_i({\cal
D}_{_{I_k}}S)\frac{\partial\,F_i}{\partial\,({\cal
D}_{_{I_k}}S)}\right]\,dS\\
\nonumber=&&\sum_{[k=0]}\partial_i({\cal
D}_{_{I_k}}\rho)\int\frac{\partial\,F_i}{\partial\,({\cal
D}_{_{I_k}}\rho)}\,dS +\sum_{[k=1]}\partial_i({\cal
D}_{_{I_k}}S)\int\frac{\partial\,F_i}{\partial\,({\cal
D}_{_{I_k}}S)}\,dS+(\partial_i\,S)\,\int\frac{\partial\,F_i}{\partial\,S}\,dS\\
\nonumber =&&\sum_{[k=0]}\partial_i({\cal
D}_{_{I_k}}\rho)\frac{\partial}{\partial\,({\cal
D}_{_{I_k}}\rho)}\int\,F_i\,dS+\sum_{[k=1]}\partial_i({\cal
D}_{_{I_k}}S) \frac{\partial}{\partial\,({\cal
D}_{_{I_k}}S)}\int\,F_i\,dS+(F_i+C_i)\,\partial_i\,S\\
=&&{\bfm\nabla}\cdot\int{\bfm F}\,dS+{\bfm C}\cdot{\bfm\nabla}S \
,\label{b5}
\end{eqnarray}
with $C_i$ integration constants not depending on $S$. Eq.
(\ref{b5}) coincides with Eq. (\ref{sol}) for ${\bfm G}=\int{\bfm
F}\,dS$ and $\overline{U}={\bfm C}\cdot{\bfm\nabla}S$.

\hspace {155mm}$\Box$


\vfill\eject
\end{document}